\documentclass[aps,amsmath,amssymb,reprint]{revtex4-1}

\usepackage{graphicx}% Include figure files
\usepackage{dcolumn}% Align table columns on decimal point
\usepackage{bm}% bold math

\usepackage[utf8]{inputenc}
\usepackage[T1]{fontenc}
\usepackage{mathptmx}

\usepackage{siunitx}
\usepackage{color}

\usepackage{tikz}
\usetikzlibrary{shapes,arrows}
\usetikzlibrary{spy,decorations.fractals,shapes.misc}
\usetikzlibrary{decorations.markings}
\usepackage{pgfplots}
\usepgfplotslibrary{groupplots}
\usepgfplotslibrary{colormaps}
\usepgfplotslibrary{colorbrewer}
\pgfplotsset{cycle list/Dark2}
\pgfplotsset{compat=newest,/pgf/number format/.cd,1000 sep={}}
\pgfplotsset{filter discard warning=false}
\usepackage[caption=false]{subfig}
\usepgfplotslibrary{colormaps}
\usepgfplotslibrary{colorbrewer}
\pgfplotsset{cycle list/Dark2}
\usepackage[capitalise]{cleveref}
\begin{document}

\title[Exponential BGK Integrator]{Exponential BGK Integrator for Multi-Scale Particle-Based Kinetic Simulations} 

\author{M. Pfeiffer}
\thanks{Corresponding author.}
 \email{mpfeiffer@irs.uni-stuttgart.de}
 \affiliation{%
   Institute of Space Systems, University of Stuttgart, Pfaffenwaldring 29, 70569
   Stuttgart, Germany
 }%

\author{F. Garmirian}
 \email{garmirianf@irs.uni-stuttgart.de}
 \affiliation{%
   Institute of Space Systems, University of Stuttgart, Pfaffenwaldring 29, 70569
   Stuttgart, Germany
 }%
 
 \author{M. H. Gorji}
 \email{Mohammadhossein.Gorji@empa.ch}
 \affiliation{%
Laboratory of Multiscale Studies in Building Physics, Empa, Swiss Federal Laboratories for Materials Science and Technology, D\"{u}bendorf, Switzerlandy
 }%

\date{\today}% It is always \today, today,
             %  but any date may be explicitly specified

\begin{abstract}
Despite the development of an extensive toolbox of multi-scale rarefied flow simulators, such simulations remain challenging due to the significant disparity of collisional and macroscopic spatio-temporal scales. Our study offers a novel and consistent numerical scheme for a coupled treatment of particles advection and collision governed by the BGK evolution, honouring positivity of the velocity distribution. Our method shares its framework, in spirit, with the unified gas kinetic class of multi-scale schemes. Yet  it provides attractive features for particle-based stochastic simulations, readily implementable to existing direct simulation Monte-Carlo codes. We demonstrate accuracy and performance of the devised scheme for prototypic gas flows, over a wide-range of rarefaction parameters. Due to the resulting robustness and flexibility of the devised exponential BGK integrator, the scheme paves the way towards more affordable simulations of large- and multi-scale rarefied gas phenomena. 
\end{abstract}

\maketitle

\section{Introduction}
The direct simulation Monte Carlo (DSMC) method has become the standard approach for simulations of rarefied non-equilibrium gas flows\cite{Bird1994}. 
It is built on the main idea of employing discrete particle collisions in order to statistically mimic the molecular collisions. An important success element of DSMC rests on the particle-based treatment of the phase-space, which allows rather straightforward integration of the inner energy modes, chemical reactions as well as boundary interactions. 
Despite the established accuracy of DSMC across the whole rarefaction regimes, it comes with a drawback of dense operations near the continuum as the mean free path and the collision frequency have to be resolved. As a consequence, it is still challenging to simulate flows with large variation of the Knudsen number, covering both continuum and rarefied phenomena. To overcome this central issue, different approaches have been proposed. The most direct solution is to couple continuum solvers based on the Navier-Stokes equations with DSMC \cite{hash1995hybrid,carlson2004hybrid}. However since the two solvers operate at different levels of the flow description, i.e. macroscopic variables in the Navier-Stokes and particle-based probabilities in DSMC, various problems arise which limit the generality and robustness of such approaches.

Another possibility is offered by multi-scale methods based on the discretisation of the phase-space, as provided e.g. in Discrete-Velocity-Methods (DVMs)\cite{mieussens2000discrete}. Along DVM, recently there have been several developments such as UGKS\cite{xu2010unified,chen2015comparative} and DUKGS\cite{guo2013discrete}, where Bhatnagar-Gross-Krook (BGK) approximations of the collision term are computed efficiently. Due to the deterministic treatment of the kinetic problem, these methods provide noise-free solutions, which make them well suited for low-Mach flows. However this comes with the price of the velocity-space discretisation, which might render these methods inefficient for high-Mach non-equilibrium flows with significantly extended velocity domains.

An interesting remedy to circumvent this problem is provided by the UGKWP method\cite{liu2020unified}. Here the non-equilibrium part of the velocity distribution is represented by particles, to reduce the corresponding cost of the velocity discretisation. Furthermore, such hybrid representation allows for noise-free solutions in the hydrodynamic limit, as particles are only employed in the non-equilibrium portions of the flow. Nevertheless, a mixed particle-DVM treatment of the distribution leads to implementation challenges, especially if the approach is to be integrated into the existing mature DSMC solvers.  

At a different front, quite a few particle-based BGK\cite{Pfeiffer2018,zhang2019particle,pfeiffer2019evaluation}  and Fokker-Planck\cite{gorji2014efficient,gorji2021entropic,mathiaud2016fokker} (FP) methods came to 
the fore in recent years, as they could simply be coupled with DSMC while also being efficient at moderate/low Knudsen flows. On the BGK side, especially the particle-based ellipsoidal statistical BGK\cite{gallis2011investigation,gallis2000application}
(ES-BGK) and the Shakhov BGK methods were investigated\cite{Pfeiffer2018,fei2020benchmark}. For the FP models, the main work was done on the Entropic FP model (EFP) \cite{gorji2021entropic}, cubic-FP model\cite{gorji2011fokker}, the ellipsoidal statistical FP (ESFP) model\cite{mathiaud2016fokker} and the model of \citet{bogomolov2009fokker}. However while the proposed schemes might reduce the cost of dense collisions in the continuum regime, they still need fine resolutions. This is due to the typical first-order treatment of particles evolution which is implied by splitting between free flight and collision/relaxation sub-time-steps.

In this study, we devise a particle-based multi-scale BGK solver. We remain in the realm of pure particle representation of the distribution with identical positive statistical weights. This avoids implementation overheads of integrating the scheme with existing DSMC solvers.  Moreover, we anticipate straightforward extensions to more complex non-equilibrium phenomena (e.g. chemical reactions). The multi-scale capability of the scheme is achieved by coupled position-velocity integration of the particles evolution, which is discussed in the following, after brief review of the governing theory. 
% The USP-BGK method\cite{fei2020unified}, which has already been successfully coupled at DSMC, is particularly worth mentioning here. However, this will be discussed in more detail later.

\section{Theory}

The Boltzmann equation describes a monatomic gas flow with the corresponding distribution function 
$f=f(\mathbf x, \mathbf v, t)$ at position $\mathbf x$ and velocity $\mathbf v$
\begin{equation}
\frac{\partial f}{\partial t} + \mathbf v \cdot \frac{\partial f}{\partial \mathbf x} = \left.\frac{\delta f}{\delta t}\right|_\mathrm{coll},
\end{equation}
where external forces are neglected and $\left.\delta f/\delta t\right|_\mathrm{coll}$ follows the Boltzmann collision integral
\begin{equation}
\left.\frac{\partial f}{\partial t}\right|_{\mathrm{coll}}=\int_{\mathbb{R}^3}\int_{S^2} B
\left[f(\mathbf v')f(\mathbf v_*')-f(\mathbf v)f(\mathbf v_*)\right]\, d\mathbf n \, d\mathbf v_*.
\end{equation}
Here, $S^2\subset\mathbb{R}^3$ is the unit sphere, $\mathbf n$ is the unit vector of the scattered velocities, $B$ is the collision
kernel and the superscript $'$ denotes post-collision velocities. The multiple integration of this non-linear collision term, besides high-dimensionality of the solution domain make the Boltzmann collision integral computationally complex. For this reason, the DSMC method reduces
 the collision integral to the Monte-Carlo sampling of  collision events among random particles representing the gas flow. Either through direct solvers or by random particles, the mean free path as well as the collision frequency must be resolved, leading to a significant computational effort for low Knudsen number flows. 

\subsection{BGK Approximation}

The BGK model approximates the collision term by a non-linear relaxation form, where the distribution $f$ relaxes towards a target $f^t$ 
\begin{equation}
\left.\frac{\partial f}{\partial t}\right|_{Coll}=\Omega = \nu\left(f^t-f\right)
\label{eq:bgkmain}
\end{equation}
with a 
certain relaxation frequency $\nu$.
It is assumed, in the original BGK model, that the target  distribution is the Maxwellian 
\begin{equation}
f^M=n\left(\frac{m}{2\pi k_B T}\right)^{3/2} \exp\left[-\frac{m\mathbf c \cdot \mathbf c}{2k_B T}\right],
\label{eq:maxwelldist}
\end{equation}
with the number density $n$, the molecular mass $m$, temperature $T$, the Boltzmann constant $k_B$ and the thermal  velocity $\mathbf c=\mathbf v -\mathbf u$ with the average flow velocity $\mathbf u$ \cite[]{bhatnagar1954model}.
The relaxation frequency
gives rise to the viscosity 
\begin{equation}
\mu=\frac{nk_BT}{\nu}.
\end{equation}
The Maxwellian distribution, as the target, leads to the fixed Prandtl number of $\textrm{Pr}=\mu c_P/K=1$, whereas the Prandtl number for monatomic gases is close to $2/3$ \cite{vincenti1965introduction}. To overcome this problem,
several extensions of the BGK model were introduced in the past. Some of these models transform the target distribution function
e.g. the ellipsoidal statistical BGK model\cite{holway1966new} or the Shakhov BGK model\cite{shakhov1968generalization}, 
while others modify the relaxation frequency from a constant to a function of the microscopic velocities as described by \citet{struchtrup1997bgk}. 

\subsection{Particle based BGK solver}
While the BGK relaxation has a much simpler construct compared to the Boltzmann collision integral, still stiff relaxation might be encountered if explicit time integrations are applied. %One main problem by solving the BGK equation is the stiff relaxation time $\tau=1/\nu$ which has to be resolved using explicit time integrations. 
%Furthermore, the equation should be integrated with a second-order method, so that the method still represents the correct viscosity of the gas for time steps greater than the relaxation time, i.e. converges asymptotically to the Navier-Stokes equations\#. 
Furthermore, to obtain asymptotical convergence to the Navier-Stokes equations, the advection (particle movement) and the relaxation term should be treated in a coupled way, as discussed by  \citet{xu2010unified,guo2013discrete}.
%As discussed in \#, this also means that the advection (particle movement) and the relaxation term should be treated in a coupled way. 

In the stochastic particle BGK method (SP-BGK), the time integration of Eq.~ \eqref{eq:bgkmain} is typically done by assuming a constant target distribution\cite{gallis2011investigation,gallis2000application}
\begin{equation}
f(\mathbf v, \mathbf x, t+\Delta t) = e^{-\nu\Delta t}f(\mathbf v, \mathbf x, t) + (1-e^{-\nu\Delta t})f^t(\mathbf v, \mathbf x, t).
\label{eq:etd1}
\end{equation}
The advection part follows a first-order operator-splitting which means that the particles of $f(\mathbf v, \mathbf x, t+\Delta t) $ are moved along their trajectory for a full time step $\Delta t$ to reach $f(\mathbf v, \mathbf x+\mathbf v \Delta t, t+\Delta t) $.
This first-order time integration has some advantages for particle methods. For $\nu\Delta t \ll 1$, the time integration Eq.~\eqref{eq:etd1} recovers the forward Euler method 
\begin{equation}
f(\mathbf v, \mathbf x, t+\Delta t) = (1-\nu\Delta t)f(\mathbf v, \mathbf x, t) + \nu\Delta t f^t(\mathbf v, \mathbf x, t).
\label{eq:euler1}
\end{equation}
Yet in contrast to the forward Euler Eq.~\eqref{eq:euler1}, the prefactors of $f(\mathbf v, \mathbf x, t)$ and $f^t$ are always positive in Eq.~\eqref{eq:etd1}, also when $\nu\Delta t > 1$. In the stochastic particle method context, Eq.~\eqref{eq:etd1} can be easily realised if each particle within a cell gets a new velocity sampled from $f^t$ with the well-posed probability $(1-e^{-\nu\Delta t})$.
For the Euler forward method, however, similar realisations only work as long as $\nu\Delta t$ remains below 1. Otherwise, particles with negative weights have to be introduced in order to construct the desired distribution function, as the prefactor of $f(\mathbf v, \mathbf x, t)$ becomes negative in Eq. \eqref{eq:euler1}. While $\nu\Delta t > 1$ may not be relevant for the first-order explicit time integration anyhow, the unconditional positivity of the distribution function reveals the advantage of  Eq.~\eqref{eq:etd1} over Eq.~\eqref{eq:euler1} for particle methods, both in terms of the robustness as well as the implementation.

To improve the explicit time integration to a second-order method coupling advection and relaxation, we follow UGKS\cite{xu2010unified,chen2015comparative} and DUGKS\cite{guo2013discrete}. The Crank-Nicolson method is used for the time integration
\begin{eqnarray}
&f(\mathbf v, \mathbf x+\mathbf v \Delta t, t+\Delta t) = f(\mathbf v, \mathbf x, t) \nonumber\\
& + \frac{\Delta t}{2}\left(\Omega(\mathbf v, \mathbf x+\mathbf v \Delta t, t+\Delta t) + \Omega(\mathbf v, \mathbf x, t) \right).
\label{eq:CN}
\end{eqnarray}
Next, let us introduce two additional distributions
\begin{eqnarray}
\tilde{f} &= f - \frac{\Delta t}{2}\Omega = \frac{2\tau+\Delta t}{2\tau}f-\frac{\Delta t}{2\tau}f^t\\ 
\textrm{and} \ \ \ \ \ \hat{f} &= f + \frac{\Delta t}{2}\Omega = \frac{2\tau-\Delta t}{2\tau}f+\frac{\Delta t}{2\tau}f^t.
\label{eq:twoCN}
\end{eqnarray}
After inserting the above-introduced distributions into  Eq.~\eqref{eq:CN}, one sees that $\tilde{f}(\mathbf v, \mathbf x+\mathbf v \Delta t, t+\Delta t)$ is nothing but the particles of $\hat{f}(\mathbf v, \mathbf x, t)$ which are moved along their trajectories for $\Delta t$, i.e.
\begin{equation}
\tilde{f}(\mathbf v, \mathbf x+\mathbf v \Delta t, t+\Delta t)=\hat{f}(\mathbf v, \mathbf x, t) .
\end{equation}
This approach provides an implicit integration of the coupled advection and relaxation, suitable in theory also for stochastic particle methods.
Note that, in the case of DVM (including UGKS and DUGKS), some additional work has to be done for the flux reconstruction. 

Nevertheless, for stochastic particle methods the problem of negative pre-factors for large $\Delta t$ arises again, so that additional particles with partly negative weighting factors would have to be added to construct the distribution functions. A practical solution was provided by \citet{fei2020unified,fei2021}, where an additional collision term was inserted in which the current distribution function is approximated by a Grad-13 approximation. This allows the advection and the relaxation process to be solved together. The choice of the Grad-13 distribution also ensures that the Navier-Stokes limit is asymptotically preserved. Finally, a multi-scale parameter was defined, allowing a smooth transition between the normal SP-BGK method and the one subject to the Grad-13 approximation. 

Even though a good performance of the SP-BGK method regularized by Grad-13 approximation has been demonstrated, it is still desirable to construct a rigorous SP-BGK algorithm with overarching properties of the DSMC method, without the need of introducing auxiliary approximations to deal with the negative pre-factors.  In the following, we construct such second-order particle method without making additional approximations on the distribution functions. Thus, the introduction of an additional multi-scale parameter is omitted and the construction would come consistent with the BGK equation, in a fashion similar to UGKS and DUGKS.

\subsection{Exponential Time Differencing of the BGK equation}
The exponential time differencing has advantages especially in dealing with ordinary differential equations that contain a stiff linear term \cite{cox2002exponential}
\begin{equation}
\dot{u} = cu + F(u,t),
\label{eq:expintexemp}
\end{equation}
where $c$ is a constant that leads to the stiff system, and $F$ is the non-linear part. The main idea is to integrate the stiff term in the exact form, and then approximate the nonlinear term numerically (typically explicitly), as following
\begin{equation}
u(t+\Delta t) = u(t)e^{c\Delta t} + e^{c\Delta t}\int_0^{\Delta t}e^{-cs} F(u(t+s),t+s)\,ds
\label{eq:expintsol}
\end{equation}
which is exact up to the point where the occurring integral is approximated.
If we now look at the relaxation term of the BGK equation in a form corresponding to Eq.~\eqref{eq:expintexemp}
\begin{equation}
\partial f/\partial t = -\nu f + \nu f^t,
\label{eq:bgkexp1}
\end{equation}
we face an additional problem  that the non-linear part $\nu f^t$ is also stiff (since it scales with $\nu$). Following Eq.~\eqref{eq:expintsol}, we get
\begin{equation}
f(t+\Delta t) = f(t)e^{-\nu \Delta t} + e^{-\nu \Delta t}\int_0^{\Delta t}e^{\nu s} \nu f^t(t+s)\,ds.
\label{eq:exactEI}
\end{equation} 
Now in order to treat the stiffness of the non-linear term, an implicit integration, such as the Crank-Nicolson method similar to the DUGKS method, can be carried out. In the following, we carry out this procedure and it is shown that it has various advantages for particle methods, since the prefactors mentioned always remain positive.
%What is interesting at this point is that for the simplest approximation with the assumption $f^t(t+s)=const.=f^t(t)$, the solution results from \eqref{eq:etd1} for the normal SP-BGK. Due to this approximation, the time intergration of  \eqref{eq:etd1} is first-order accurate.
\subsubsection{Linear approximation}
Let $f_{t+s}^t$ and $f_t^t$ represent the numerical approximations of the target distribution at time steps $t+s$ and $t$, respectively. To construct a second-order scheme, we adopt the following linear approximation for the target distribution
\begin{equation}
f^t_{t+s}= f^t_t+ \frac{s}{\Delta t}(f^t_{t+\Delta t}-f^t_t)
\label{eq:linearapprox}
\end{equation}
leading to a Crank-Nicolson type scheme.
Using this approximation, the integration of Eq.~\eqref{eq:exactEI} yields to
\begin{eqnarray}
f_{t+\Delta t} &= f_t e^{-\nu \Delta t} + e^{-\nu \Delta t}\left[ f^t_t\left( \frac{e^{\nu \Delta t}}{\nu \Delta t} - 1 -\frac{1}{\nu \Delta t}\right)\right. \nonumber \\
&\left.+f^t_{t+\Delta t}\left( e^{\nu \Delta t} + \frac{1}{\nu \Delta t} - \frac{e^{\nu \Delta t}}{\nu \Delta t}\right)\right],
\label{eq:linexp}
\end{eqnarray}
where $f_{t+\Delta t}$ and $f_t$ denote the numerical approximations of $f(t+\Delta t)$ and $f(t)$, respectively.
After a few rearrangements and taking particle movement into account, Eq.~\eqref{eq:linexp} leads to the following integration scheme
\begin{eqnarray}
f_{t+\Delta t}&(\mathbf v, \mathbf x+\mathbf v \Delta t) = f_t(\mathbf v, \mathbf x)e^{-\nu \Delta t} + \left(1-e^{-\nu \Delta t}\right)\times \nonumber 
\\ 
&\left[Af^t_t(\mathbf v, \mathbf x) + Bf^t_{t+\Delta t}(\mathbf v, \mathbf x+\mathbf v \Delta t) \right],
\label{eq:linfin}
\end{eqnarray}
with 
\begin{eqnarray}
A &=& \frac{1}{\nu \Delta t} - \frac{e^{-\nu\Delta t}}{1-e^{-\nu\Delta t}}\\
\textrm{and} \ \ \ \ B &=& \frac{1}{1-e^{-\nu\Delta t}}-\frac{1}{\nu\Delta t}.
\end{eqnarray}
Equation \eqref{eq:linfin} already contains the simple form of the first order SP-BGK method \eqref{eq:etd1}. Furthermore the prefactors $A\in[0,0.5]$ and 
$B\in[0.5,1]$ are non-negative. 
Following \citet{xu2010unified,guo2013discrete}, two additional distributions can be introduced
\begin{eqnarray}
\hat{f}  &=&  f_t e^{-\nu \Delta t} + (1-e^{-\nu \Delta t})Af_t^t\label{eq:linexpa}\\
\textrm{and} \ \ \ \ \tilde{f} &=& f_t - (1-e^{-\nu \Delta t})Bf_t^t, \label{eq:linexpb}
\end{eqnarray}
whereby once more 
\begin{equation}
\tilde{f}(\mathbf v, \mathbf x+\mathbf v \Delta t, t+\Delta t) = \hat{f}(\mathbf v, \mathbf x, t)
\end{equation}
is honoured.
Therefore, by moving the particles of $\hat{f}(\mathbf v, \mathbf x,t)$ along the trajectories $\tilde{f}(\mathbf v, \mathbf x+\mathbf v \Delta t, t+\Delta t)$, the advection and relaxation processes are coupled and the time integration remains second-order accurate. Starting from $f_t(\mathbf v, \mathbf x)$, $\hat{f}(\mathbf v, \mathbf x, t)$ can be constructed from Eq.~\eqref{eq:linexpa}. Afterwards, $\tilde{f}(\mathbf v, \mathbf x+\mathbf v \Delta t, t+\Delta t)$ is realised by moving the particles from $\hat{f}(\mathbf v, \mathbf x, t)$, and finally $f_{t+\Delta t}(\mathbf v, \mathbf x+\mathbf v \Delta t)$ is constructed via
\begin{equation}
f_t = \tilde{f}  + (1-e^{-\nu \Delta t})Bf_t^t.
\label{eq:frecon}
\end{equation}
While all prefactors remain non-negative, a re-normalisation has to be introduced as the prefactors of Eqs. \eqref{eq:linexpa} and \eqref{eq:frecon} do not sum up to unity. To avoid deleting and creating particles, the introduced distribution functions are normalized by the factor
\begin{equation}
\gamma = e^{-\nu \Delta t} + (1-e^{-\nu \Delta t})A = \frac{1-e^{-\nu \Delta t}}{\nu \Delta t}
\end{equation}
which results in new distributions $\hat{f}^*=\hat{f}/\gamma$ and $\tilde{f}^*=\tilde{f}/\gamma$, with $\tilde{f}^*(\mathbf v, \mathbf x+\mathbf v \Delta t, t+\Delta t) = \hat{f}^*(\mathbf v, \mathbf x, t)$, leading to
\begin{eqnarray}
\hat{f}^* &=&   \frac{e^{-\nu \Delta t}}{\gamma}f_t + \frac{1-e^{-\nu \Delta t}}{\gamma}Af_t^t \nonumber \\
      &=& \frac{\nu\Delta t e^{-\nu \Delta t}}{1-e^{-\nu\Delta t}}f_t + \left(1-\frac{\nu\Delta t e^{-\nu \Delta t}}{1-e^{-\nu\Delta t}}\right)f_t^t\label{eq:finalin}\\
\textrm{and} \ \ \ \ f_t &=& \gamma \tilde{f}^* + (1-e^{-\nu \Delta t})Bf_t^t \nonumber \\
&=& \frac{1-e^{-\nu\Delta t}}{\nu \Delta t} \tilde{f}^* + \left(1-\frac{1-e^{-\nu\Delta t}}{\nu \Delta t}\right)f_t^t\label{eq:finblin}.
\end{eqnarray}
To further simplify the time integration, one can also omit the intermediate step via $f$ and only track the additional distribution functions $\tilde{f}^*$ and $\hat{f}^*$. By substituting Eq.~\eqref{eq:finalin} into \eqref{eq:finblin}, we get
\begin{equation}
\hat{f}^* = e^{-\nu \Delta t} \tilde{f}^* + (1-e^{-\nu \Delta t})\underbrace{\left[\frac{Be^{-\nu \Delta t}}{\gamma}+ \frac{A}{\gamma}\right]}_{=1}f_t^t.
\label{eq:finallincom}
\end{equation}
This is a convenient formulation as similar to the original SP-BGK, particles have to be sampled from the target distribution with the probability $(1-e^{-\nu \Delta t})$ to construct $\hat{f}^*$ from $\tilde{f}^*$. The two most common target distributions are the ellipsoidal statistical BGK (ES-BGK) and the Shakhov model (S-BGK). The ES-BGK distribution is given as 
\begin{equation}
f^{ES}=\frac{n}{\sqrt{\det (2\pi\lambda_{ij})}}\exp\left[-\frac{1}{2}\lambda_{ij}^{-1}c_ic_j\right]
\label{eq:esbgkdist}
\end{equation}  
with the matrix 
\begin{equation}
\lambda_{ij} = \frac{k_BT}{m}\delta_{ij} + \left(1-\frac{1}{\textrm{Pr}}\right)\frac{p_{\langle ij\rangle}}{\rho}.
\end{equation}
Here, $\delta_{ij}$ is the Kronecker delta, $\textrm{Pr}$ is the Prandtl number, $\rho$ is the mass density and $p_{\langle ij \rangle}$ is the trace-less pressure tensor
\begin{equation}
p_{\langle ij \rangle}=m\int c_{\langle i}c_{j\rangle}f\,d\mathbf c.
\end{equation}
In the case of the S-BGK model, the target distribution is defined as
\begin{equation}
f^S=f^M\left[1+(1-\textrm{Pr})\frac{\mathbf c \cdot \mathbf q}{5 \rho (RT)^2}\left(\frac{\mathbf c \cdot \mathbf c}{2RT}-\frac{5}{2}\right)\right]
\end{equation}
with the heat flux vector
\begin{equation}
\mathbf q = \frac{1}{2}m\int \mathbf c (\mathbf c \cdot \mathbf c) f\,d\mathbf c.
\label{eq:heatflux}
\end{equation}
Therefore to evaluate Eq.~\eqref{eq:finallincom}, either knowledge about $p_{\langle ij \rangle}(f)$ or $\mathbf q(f)$ become necessary.
Due to the normalization of the distributions, as already extensively discussed by \citet{xu2010unified,guo2013discrete}, the mass, momentum and energy are conserved by the collision operator and can be thus determined directly from the additional distributions
\begin{eqnarray}
\rho&=&\int m f\,d\mathbf v=\int m \hat{f}^*\,d\mathbf v=\int m \tilde{f}^*\,d\mathbf v,\\
\rho\mathbf u&=&\int m \mathbf v f\,d\mathbf v=\int m  \mathbf v \hat{f}^*\,d\mathbf v=\int m \mathbf v  \tilde{f}^*\,d\mathbf v \\
\textrm{and} \ \ \ \rho\epsilon&=&\frac{3}{2}n k_BT \nonumber \\
  &=& \int \frac{m}{2} \mathbf c^2 f\,d\mathbf v=\int \frac{m}{2}  \mathbf c^2 \hat{f}^*\,d\mathbf v=\int \frac{m}{2} \mathbf c^2  \tilde{f}^*\,d\mathbf v \nonumber \\
\end{eqnarray}
with $\rho \epsilon$ being the internal energy.
Neither $p_{\langle ij \rangle}(f)$ nor $\mathbf q(f)$ would be preserved during the relaxation process, yet can be calculated directly from $\tilde{f}^*$ 
\begin{eqnarray}
p_{\langle ij \rangle}(f_t) &=& \frac{1-e^{-\nu\Delta t}}{\nu \Delta t} p_{\langle ij \rangle}(\tilde{f}^*)\label{eq:presstenscon}\\
\textrm{and} \ \ \ \mathbf q(f_t) &=& \frac{1-e^{-\nu\Delta t \textrm{Pr}}}{\nu \Delta t \textrm{Pr}} \mathbf q(\tilde{f}^*).\label{eq:heatfluxcon}
\end{eqnarray}
Following the ES-BGK target distribution, we can summarize the algorithmic steps of the devised second-order time integration, whose computatinoal complexity is almost identical to the first-order conventional SP-BGK methods:
%Thus everything is finally available for the second order time integration, whereby the computational effort is almost identical to the case of the normal first order SPBGK method, since virtually the identical things have to be done. In the following, the ESBGK target distribution function is used. This results in the following steps:
\begin{enumerate}
\item Initialize the particles in the simulation domain (similar to DSMC) method to obtain $f_t$.
\item Use Eq.~\eqref{eq:finalin} to construct $\hat{f}^*$ from $f_t$ by sampling the particles from $f_t^t$ with the probability $1-{\nu\Delta t e^{-\nu \Delta t}}/{(1-e^{-\nu\Delta t})}$.
\item Move the particles of $\hat{f}^*$ in the physical domain (and apply the boundary conditions similar to DSMC)  to construct $\tilde{f}^*(\mathbf v, \mathbf x+\mathbf v \Delta t, t+\Delta t) = \hat{f}^*(\mathbf v, \mathbf x, t)$.
\item Construct $\hat{f}^*(\mathbf v, \mathbf x+\mathbf v \Delta t, t+\Delta t)$ out of $\tilde{f}^*(\mathbf v, \mathbf x+\mathbf v \Delta t, t+\Delta t)$ using Eq.~\eqref{eq:finallincom},
by sampling the particles from $f_t^t$ with the probability $(1-e^{-\nu \Delta t})$ given the pressure tensor and heat fluxes from Eqs. \eqref{eq:presstenscon} and \eqref{eq:heatfluxcon}.
\item Repeat steps 3 and 4 for the entire duration of the simulation.
\end{enumerate}

\subsubsection{Exponential approximation}
The devised scheme can be further improved by applying exponential relaxation
\begin{equation}
f_{t+s}^t=f^t_t+ \frac{1-e^{-\nu s}}{1-e^{-\nu \Delta t}}(f^t_{t+\Delta t}-f^t_t)
\label{eq:exponapprox}
\end{equation}
instead of the linear approximation of  Eq.~\eqref{eq:linearapprox}. The two additional distributions are then modified with the prefactorss 
\begin{eqnarray}
A^{(exp)} &=& 1- \frac{1}{1-e^{-\nu\Delta t}} + \frac{\nu\Delta t e^{-\nu\Delta t}}{\left(1-e^{-\nu\Delta t}\right)^2}\\
B^{(exp)} &=&  \frac{1}{1-e^{-\nu\Delta t}} - \frac{\nu\Delta t e^{-\nu\Delta t}}{\left(1-e^{-\nu\Delta t}\right)^2}
\end{eqnarray}
resulting in a modified $\gamma^{(exp)}$ for the normalization
\begin{equation}
\gamma^{(exp)} = e^{-\nu \Delta t} + (1-e^{-\nu \Delta t})A^{(exp)} = \frac{\nu \Delta te^{-\nu \Delta t}}{1-e^{-\nu \Delta t}}.
\end{equation}
Thus the normalized additional distributions for the exponential approximation are given as
\begin{eqnarray}
\hat{f}^{*,(exp)} &=& \frac{1-e^{-\nu\Delta t}}{\nu\Delta t}f + \left(1-\frac{1-e^{-\nu\Delta t}}{\nu\Delta t}\right) f_t^t\label{eq:expapfhat}\\
\textrm{and} \ \ \ \ 
f_t &=& \frac{\nu\Delta t e^{-\nu \Delta t}}{1-e^{-\nu\Delta t}}\tilde{f}^{* ,(exp)} + \left(1-\frac{\nu\Delta t e^{-\nu \Delta t}}{1-e^{-\nu\Delta t}}\right)f_t^t\label{eq:expapftilde}.
\end{eqnarray}
Interestingly, this exponential approximation just swaps the pre-factors in Eqs.~\eqref{eq:expapfhat} and \eqref{eq:expapftilde} compared to the Eqs.~ \eqref{eq:finalin} and \eqref{eq:finblin} of the linear approach .
Therefore, this results again in 
\begin{equation}
\hat{f}^{*,(exp)} = e^{-\nu \Delta t} \tilde{f}^{*,(exp)} + (1-e^{-\nu \Delta t})f_t^t.
\label{eq:finalexpcom}
\end{equation}
The only difference that arises is the calculation of the pressure tensor and the heat flux from $\tilde{f}^{*,(exp)}$
\begin{eqnarray}
p_{\langle ij \rangle}(f_t) &=&\frac{\nu\Delta t e^{-\nu \Delta t}}{1-e^{-\nu\Delta t}} p_{\langle ij \rangle}(\tilde{f}^{*,(exp)})\label{eq:presstensconexp}\\
\textrm{and} \ \ \ \  \mathbf q(f_t) &=&\nu\Delta t \textrm{Pr}\   \frac{e^{-\nu \Delta t Pr}}{1-e^{-\nu\Delta t Pr}} \mathbf q(\tilde{f}^{*,(exp)}).\label{eq:heatfluxconexp}
\end{eqnarray}
\subsubsection{Properties of the ED-BGK method}
The behaviour of the introduced particle scheme with exponential differencing (ED-BGK) is 
analyzed in the limits of free-molecular and continuum regimes. In the following and without loss of generality, the discussion is limited to the linear approximation.
In the free molecular regime ($\tau\gg\Delta t$), Eqs.~\eqref{eq:finalin} and \eqref{eq:finblin} simplify to
\begin{equation}
\hat{f}^* = f_t,\quad f_t=\tilde{f}^*
\end{equation}
which is consistent with the collision-less limit implying $f(\mathbf v, \mathbf x+\mathbf v \Delta t, t+\Delta t)=f(\mathbf v, \mathbf x, t)$. On the other hand, for the continuum limit ($\tau\ll \Delta t$), the distribution  relaxes after a time step $\Delta t$ based Eqs.~\eqref{eq:finalin} and \eqref{eq:finblin} gives  (shown in \cref{sec:append})
\begin{eqnarray}
&f&(\mathbf v, \mathbf x+\mathbf v \Delta t, t+\Delta t)\approx f^M(\mathbf v, \mathbf x, t)\\
&-&\tau(\partial_t + \mathbf v \nabla) f^M(\mathbf v, \mathbf x, t) 
+\Delta t \partial_t f^M(\mathbf v, \mathbf x, t) 
\end{eqnarray}
which recovers the Chapman-Enskog approximation of the Navier-Stokes solution as discussed by \citet{guo2013discrete}.
\section{Implementation}
The presented BGK particle method is implemented in the PIC-DSMC code PICLas \cite{Munz2014,fasoulas2019combining} as described in details by Pfeiffer\cite{Pfeiffer2018}. The main concept of the particle BGK method especially the energy and momentum conservation, is based on the works of \citet{gallis2011investigation,gallis2000application}. The particles are moved in the physical space sorted in a computational mesh,
collide with boundaries and their macroscopic properties are sampled, all similar to DSMC. However, in contrast to the DSMC method, the collision step is replaced by a jump relaxation with the probability
\begin{equation}
P=1-\exp\left[-\nu \Delta t\right],
\label{eq:bgkrelax}
\end{equation}
according to Eq. \eqref{eq:finallincom}, towards the target distribution. $\hat{f}^*$ is constructed out of $f$ by using Eq.~\eqref{eq:finalin} or \eqref{eq:expapfhat} for the linear or the exponential approximation, respectively. Next, $\tilde{f}^*$ is constructed by moving the particles from $\hat{f}^*$ in the next time steps. The difference to the common SP-BGK method is that the pressure tensor or heat flux vector must be relaxed according to equations \eqref{eq:presstenscon} and \eqref{eq:heatfluxcon} or \eqref{eq:presstensconexp} and \eqref{eq:heatfluxconexp} depending whether the linear or the exponential approximation is used before sampling from the target distribution, since these are from $f$ and not from $\tilde{f}^*$. For the ellipsoidal statistical target distribution, only the pressure tensor, and for the Shakhov target distribution, only the heat flux vector is necessary.

To determine the correct relaxation frequency $\nu$, the well known temperature dependency of the viscosity $\mu$
\begin{equation}
\mu=\mu_{ref}\left(\frac{T}{T_{ref}}\right)^{\omega_{VHS}}
\end{equation}
is used, with the reference temperature $T_{ref}$, and $\mu_{ref}$ the reference dynamic viscosity at $T_{ref}$ \cite{burt2006evaluation},
and $\omega_{VHS}$ as the parameter of the variable hard sphere model (VHS).  
For a VHS gas, the reference dynamic viscosity can be calculated with the VHS reference diameter $d_{ref}$ of the particles
\begin{equation}
\mu_{ref}=\frac{30\sqrt{mk_BT_{ref}}}{\sqrt{\pi}4(5-2\omega_{VHS})(7-2\omega_{VHS})d_{ref}^2}.
\end{equation}

The sampling process itself as well as a detailed discussion of the possible energy and momentum conservation schemes for the particle BGK method can be found in \citet{Pfeiffer2018}.
For momentum and energy conservation, the final velocities of the particles are corrected according to
\begin{equation}
\mathbf v_i^*=\mathbf u + \alpha(\mathbf v'_i-\mathbf u'),
\label{eq:transenmom}
\end{equation}
where $\mathbf u=\sum_{i=1}^N \mathbf v_i/N$ is the bulk flow velocity before the relaxation, 
$\mathbf v'_i$ the particle velocity after the relaxation (uncorrected), and $\mathbf u'=\sum_{i=1}^N \mathbf v'_i/N$. Note that, 
$\mathbf v'_i = \mathbf v_i$, if no relaxation occurs for particle $i$. Due to
\begin{equation}
\sum_{i=1}^N(\mathbf v'_i-\mathbf u')=0, 
\end{equation} 
Equation \eqref{eq:transenmom} ensures the momentum conservation. The energy conservation is achieved by choosing $\alpha$ as
\begin{equation}
\alpha = \sqrt{\frac{T}{T'}}
\end{equation}
with $T$ the temperature before relaxation, and $T'$ after the relaxation process (in the absence of energy correction).

In order to achieve second-order accuracy in space, the values that enter the BGK operator are linearly interpolated, instead of being constant values per computational cell. A more detailed discussion, including an alternative to the linear interpolation, especially for particle methods, can be found in \citet{fei2020unified,fei2021}.
In this work, a conventional linear interpolation was used, similar to Particle-In-Cell codes, as have already been described for the PICLas code by \citet{fasoulas2019combining}, among others.

\subsection{Homogeneous relaxation}
\begin{figure}\centering
  \subfloat[Normalized shear stress]{\includegraphics{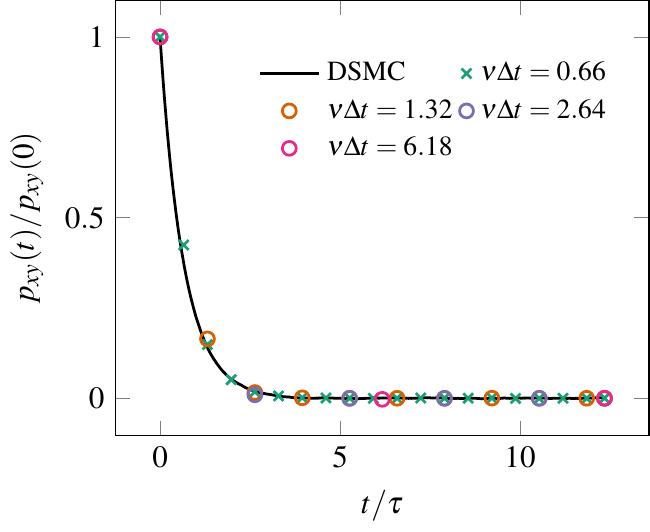}}\\
  \subfloat[Normalized heat flux]{\includegraphics{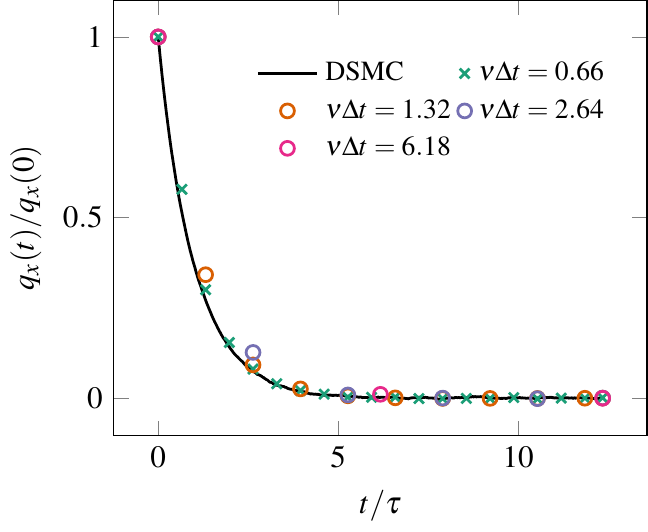}}
  \caption{Pressure tensor and heat flux relaxation with different time step sizes $\Delta t$ using the ED-SP-BGK method with the exponential approximation. The finely resolved DSMC results are provided as the reference.}\label{fig:homreledspbgkexp}
\end{figure}

The first test case, similar to \citet{fei2020unified}, considers a spatially homogeneous relaxation (in a stationary frame of reference) to examine the accuracy of the time integration scheme. In a single adiabatic cell, argon with the particle density  
$n=2.7\cdot 10^{25}\,\mathrm{m}^{-3}$ and temperature $T=273\,\mathrm{K}$ is simulated, subject to the initial condition of the 13-moment Grads distribution \cite{struchtrup2003regularization}
\begin{equation}
f^{Grad}=f^M\left[1+\frac{m^2p_{\langle ij \rangle}}{2\rho k_B^2 T^2}c_{\langle i}c_{j\rangle}-\frac{m^2q_j c_j}{\rho k_B^2 T^2}\left(1-\frac{m}{5 k_BT}\mathbf c^2\right)\right].
\end{equation} 
This allows to initialize the gas in a non-equilibrium state with stresses and heat fluxes. Each deviatoric entry of the pressure tensor was chosen according to 
$p_{\langle ij \rangle}=3nk_B T$, whereas the heat flux is set based on $q_i=n(3k_BT)^{3/2}m^{-1/2}$.
 Relaxation of the pressure tensor and the heat flux vector is analyzed for different time step sizes. The normal SP-BGK method, the proposed exponential differencing SP-BGK (ED-SP-BGK) with the linear approximation (Eq.~\eqref{eq:linearapprox}), and the exponential approximation (Eq.~\eqref{eq:exponapprox}) are employed. A finely resolved DSMC calculation is deployed as the reference.
\begin{figure}\centering
  \subfloat[Normalized shear stress]{\includegraphics{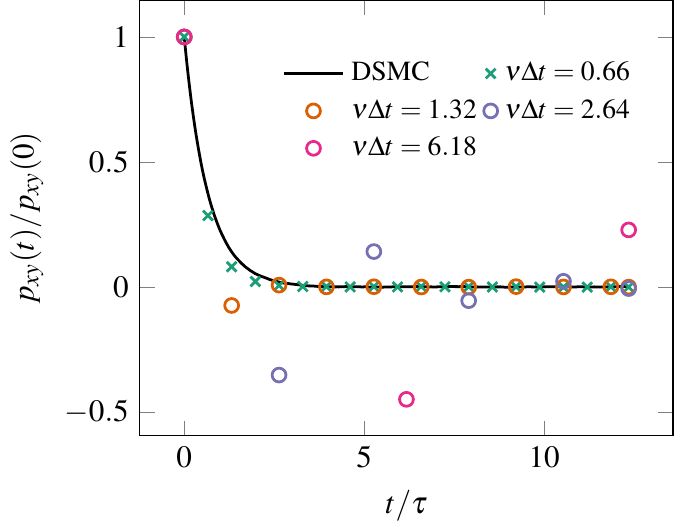}}\\
  \subfloat[Normalized heat flux]{\includegraphics{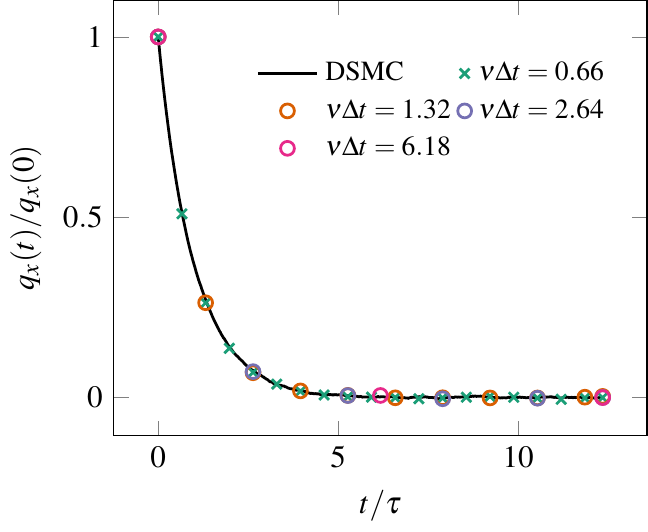}}
  \caption{Pressure tensor and heat flux relaxations with different time step sizes $\Delta t$ using SP-BGK method. The finely resolved DSMC results are provided as the reference.}\label{fig:homrelspbgk}
\end{figure}
As shown in \cref{fig:homrelspbgk}, the error in the pressure tensor for the standard SP-BGK is significant, as the target distribution is approximated as a constant per time step. 
\begin{figure}\centering
  \subfloat[Normalized shear stress]{\includegraphics{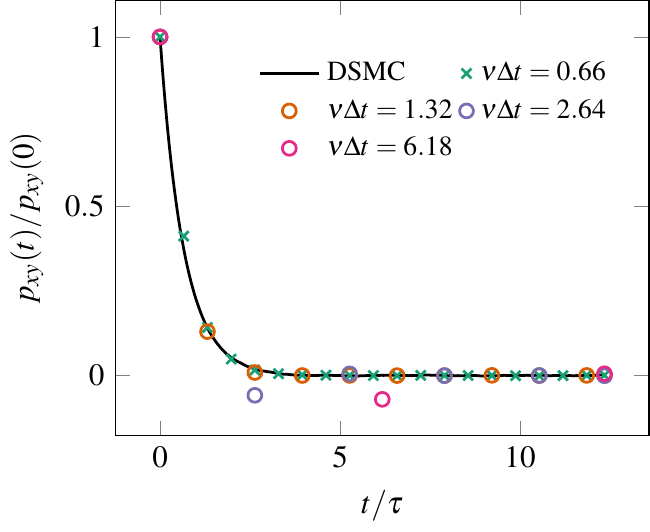}}\\
  \subfloat[Normalized heat flux]{\includegraphics{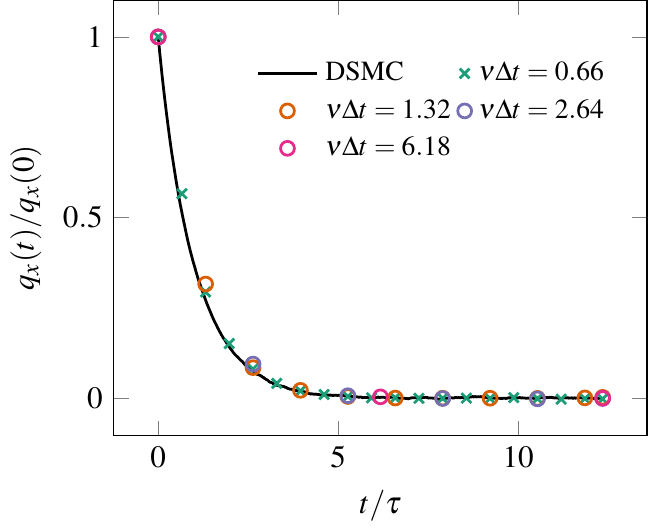}}
  \caption{Pressure tensor and heat flux relaxations with different time step sizes $\Delta t$ using the ED-SP-BGK method with the linear approximation. The finely resolved DSMC results are provided as the reference.}\label{fig:homreledspbgklin}
\end{figure}
Comparing the results depicted in \cref{fig:homreledspbgklin} and \cref{fig:homreledspbgkexp}, it can be seen that the relaxation is also correctly reproduced for time steps greater than the collision time by the devised ED-SP-BGK model. In the case of the linear approximation, the error in the relaxation of the pressure tensor is greater than the one resulting from the exponential approximation, while the heat flux is reproduced slightly better with the linear approximation.
\subsection{Sod shock tube}
The well-known Sod test case is considered in two different regimes. Argon was again used for the simulations. The physical domain has a length of 1~m. The starting temperature in the entire simulation domain is 273~K. However, the particle density differs in the left $n_l,\, x\in[0,0.5]$ and the right $n_r,\, x\in[0.5,1]$ sub-domains. 
For the rarefied case, the densities $n_l=1.508E19\,\mathrm{m^{-3}}$ and $n_r=1.885E18\,\mathrm{m^{-3}}$ were chosen. This results in a Knudsen number of about $Kn\approx 0.1$. In the continuum case, the imposed densities are $n_l=1.508E21\,\mathrm{m^{-3}}$ and $n_r=1.885E20\,\mathrm{m^{-3}}$ resulting in a Knudsen number $Kn\approx0.001$. The reference solutions were generated with DSMC. 

\begin{figure}\centering
  \subfloat[Number density]{\includegraphics{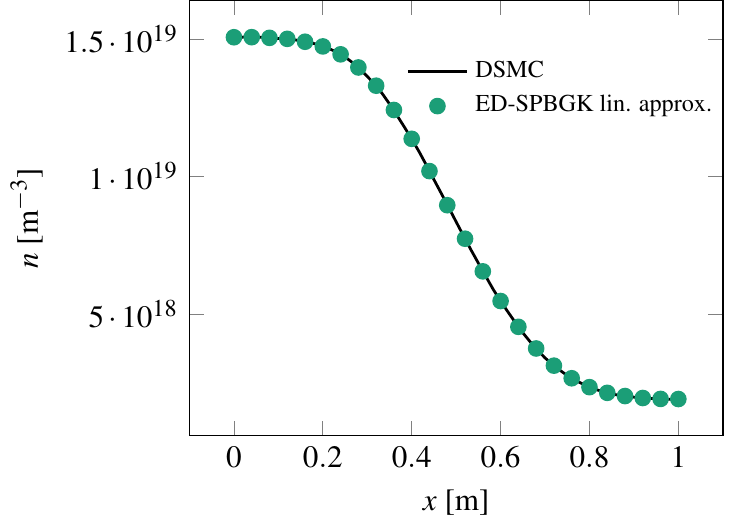}}\\
  \subfloat[Temperature]{\includegraphics{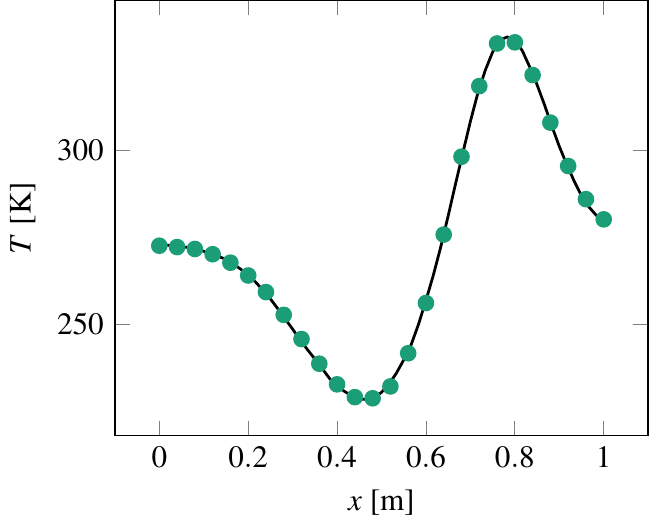}}\\
  \subfloat[Velocity]{\includegraphics{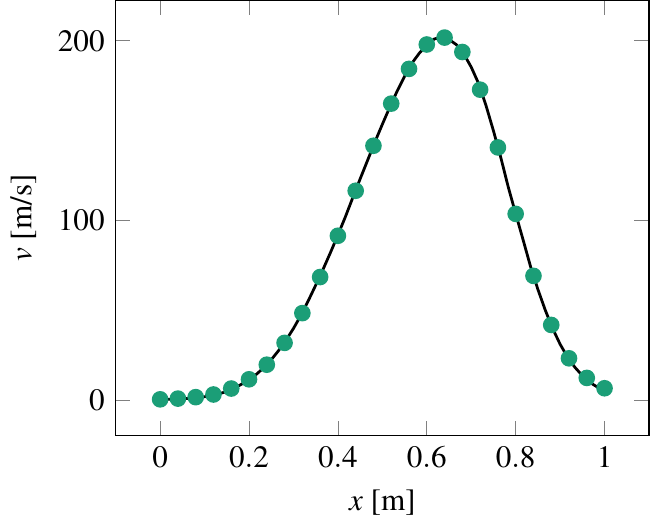}}
  \caption{Rarefied Sod shock tube testcase. ED-SP-BGK and reference DSMC results were computed using same spatio-temporal resolution.}\label{fig:SODrare}
\end{figure}

In the rarefied case, a computational mesh with 60 grid cells was used for both DSMC and ED-SP-BGK simulations. Also the time step $\Delta t= 3E-5\,\mathrm{s}$ was the same for both methods, resulting in $\nu\Delta t\approx0.035$ for the ED-SP-BGK method.
As depicted in \cref{fig:SODrare}, the DSMC result is very well matched with the ED-SP-BGK method.

\begin{figure}\centering
  \subfloat[Number density]{\includegraphics{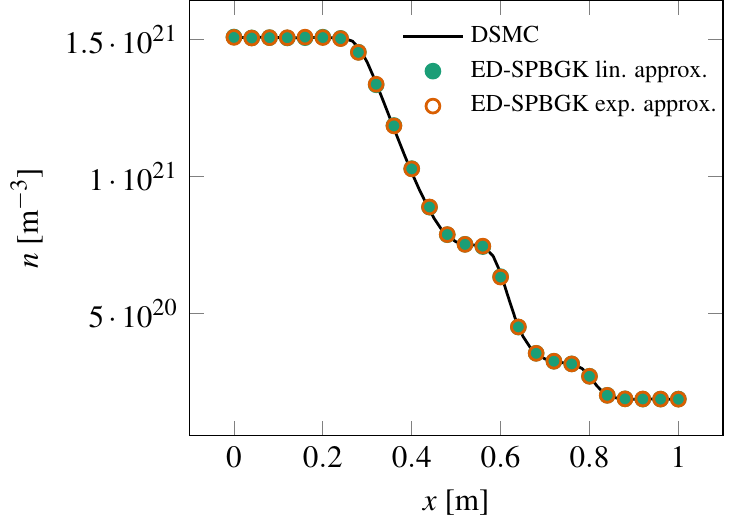}}\\
  \subfloat[Temperature]{\includegraphics{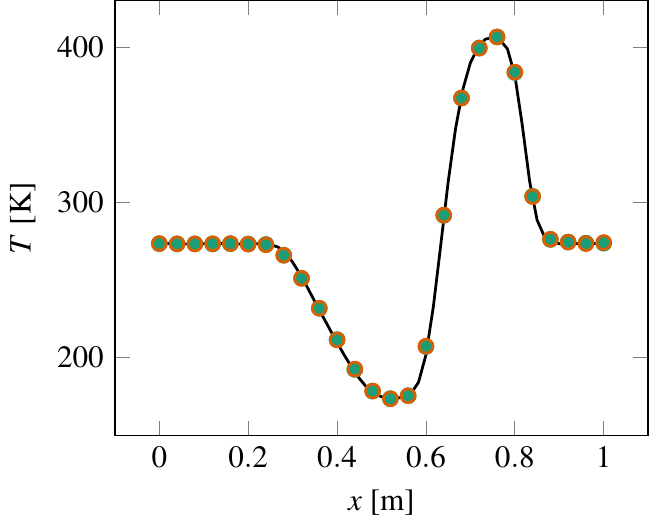}}\\
  \subfloat[Velocity]{\includegraphics{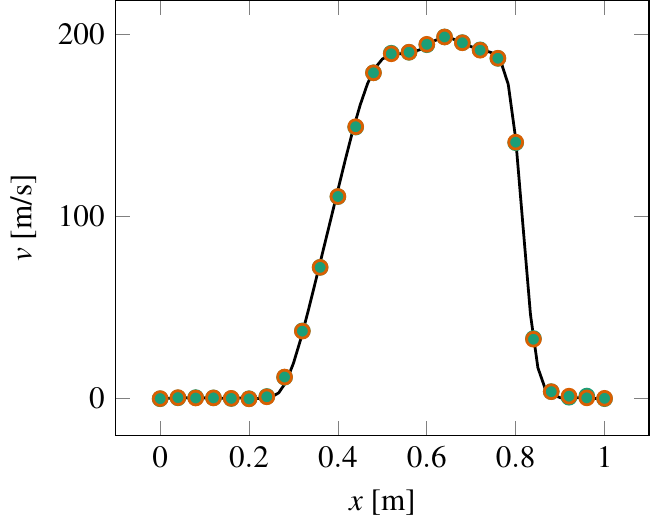}}
  \caption{Continuum Sod shock tube testcase. ED-SP-BGK and reference DSMC results were computed, where much coarser resolution has been employed for ED-SP-BGK.}\label{fig:SODhighdens}
\end{figure}

In the case of the continuum flow, while spatio-temporal discretization of DSMC was adjusted to resolve the collisional scales, the mesh similar to the rarefied setting was used for ED-SP-BGK. Therefore, an adaptive subcell method besides the time step size of $\Delta t_{DSMC}= 2E-7\,\mathrm{s}$ were used to resolve the mean free path and the mean collision time for the DSMC method. The ED-SP-BGK simulations were performed with the same time step as in the rarefied simulation. Thus, a factor of $\nu\Delta t\approx3.5$ was obtained. As can be seen in \cref{fig:SODhighdens}, the ED-SP-BGK method achieves similar number density, mean velocity and temperature with significantly coarser discretization. Both the linear and exponential approximations were tested, yielding negligible differences. For the next test cases, therefore, only the linear approximation was used.
\subsection{Couette flow}
The next test case is a steady planar Couette flow of argon, confined between the two parallel
plates which move in opposite directions with the velocity $v_{wall}=\pm 500\,\mathrm{m/s}$. The initial temperature of the gas as well as the wall temperature are $T=273\,\mathrm{K}$ leading to a supersonic Couette flow. The Couette flow was carried out in three different Knudsen number ranges. 
The conditions are shown in \cref{tab:couettecond}. In the table, the reference time step is also given, for which the solution of the standard SP-BGK method becomes identical to the proposed ED-SP-BGK method. In order to investigate solely the influence of the time integration on the performance of the two methods, identical spatial interpolation was performed in both SP-BGK and ED-SP-BGK. \\ \ \\
An important feature of wall bounded gas flows is the gas-boundary interaction. In stochastic particle methods, the time and location where a particle hits the boundary are estimated using linear interpolation. This has been shown to reset the scheme back to the first order \cite{higham2013mean,gobet2004exact}, close to the boundaries. Consistent high-order treatment of boundaries normally involve extensive modifications in the underlying stochastic process (e.g. see walk on sphere scheme for the case of random walks \cite{sabelfeld2016random,muller1955some}) Such treatments go beyond the scope of this study, yet it should be noted that a certain reduction in the order of time integration will be observed in following scenarios where boundary interactions become dominant.   
\begin{table}
\caption{Start conditions of Couette flow simulations.}\label{tab:couettecond}
  \begin{ruledtabular}\renewcommand*{\arraystretch}{1.4}
    \begin{tabular}{ccccc}
        &  \# computational cells  & number density [m$^{-3}$] & Kn & $\Delta t_{ref}$ [s] \\
      \hline
      Case 1 & 25 & $1.37E19$ & 0.1 & $1E-5$ \\
      Case 2 & 100 & $1.37E20$ & 0.01 & $5E-6$ \\
      Case 3 & 500 & $1.37E21$ & 0.001 & $6.25E-7$ \\
    \end{tabular}
  \end{ruledtabular}
\end{table}
\subsubsection{Kn=0.1}
\begin{figure}\centering
  \subfloat[ED-SPBGK]{\includegraphics{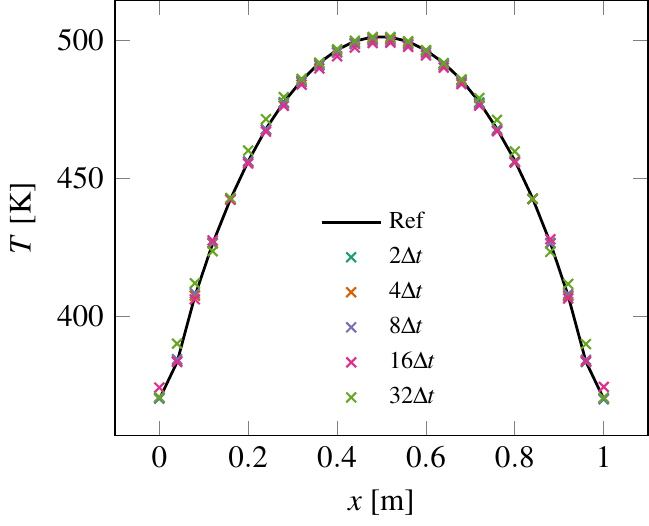}}\\
  \subfloat[SPBGK]{\includegraphics{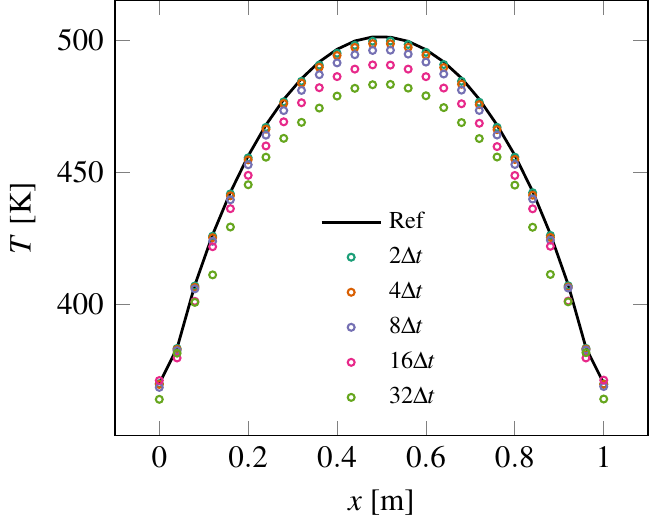}}
  \caption{Comparison of temperature using ED-SP0BGK and SP-BGK as well as different time steps $\Delta t$ for Couette flow with $Kn=0.1$.}\label{fig:couettekn01}
\end{figure}

\begin{figure}\centering
  \subfloat[ED-SPBGK]{\includegraphics{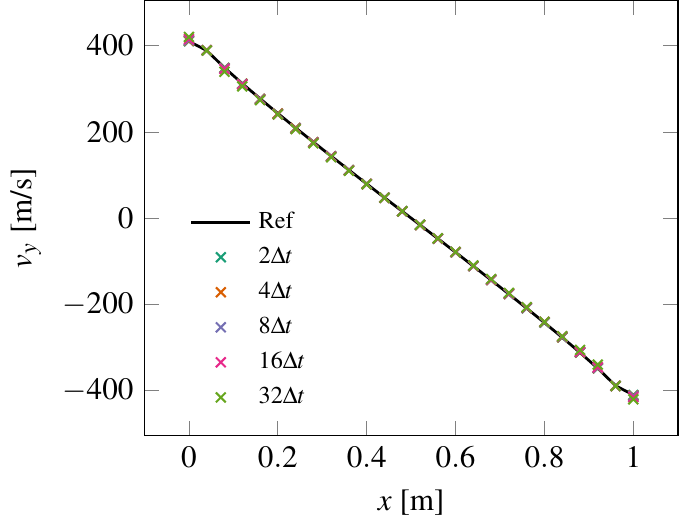}}\\
  \subfloat[SPBGK]{\includegraphics{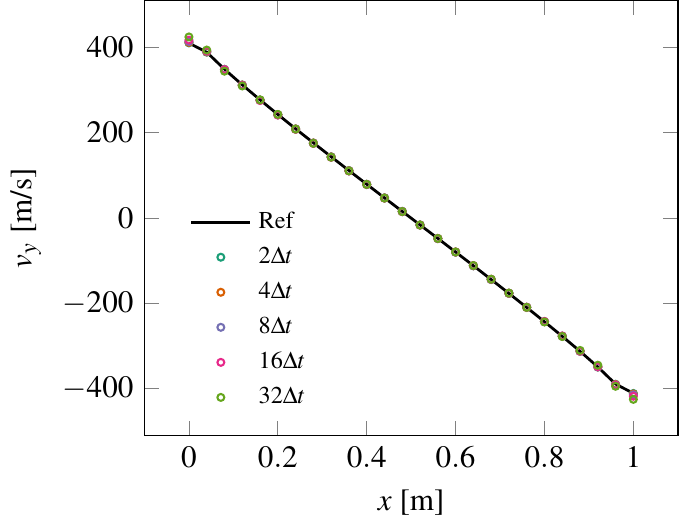}}
  \caption{Comparison of velocity $v_y$ using ED-SP-BGK and SP-BGK as well as different time steps $\Delta t$ for Couette flow with $Kn=0.1$.}\label{fig:couettekn01velo}
\end{figure}

The first Couette flow is in the rarefied range with $Kn=0.1$. \cref{fig:couettekn01} shows results of ED-SP-BGK and SP-BGK with different time step sizes. For the coarsest case of $32\Delta t$, a factor of maximum $\nu\Delta t\approx0.8$ is obtained in the simulation region. We observe that the ED-SP-BGK method achieves significantly better results compared to the SP-BGK method, even in the rarefied range where $\nu\Delta < t$.
In \cref{fig:couettekn01velo}, the velocity in y-direction $v_y$ is depicted. Here, both methods match the reference solution quite well for the different time step sizes.
\subsubsection{Kn=0.01}
\begin{figure}\centering
  \subfloat[ED-SPBGK]{\includegraphics{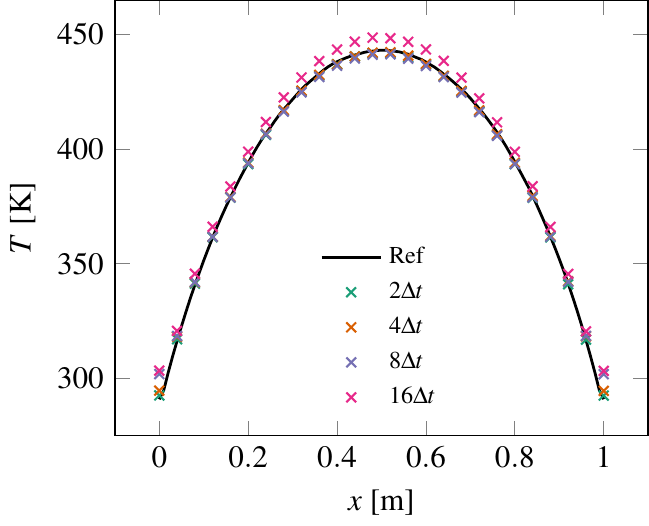}}\\
  \subfloat[SPBGK]{\includegraphics{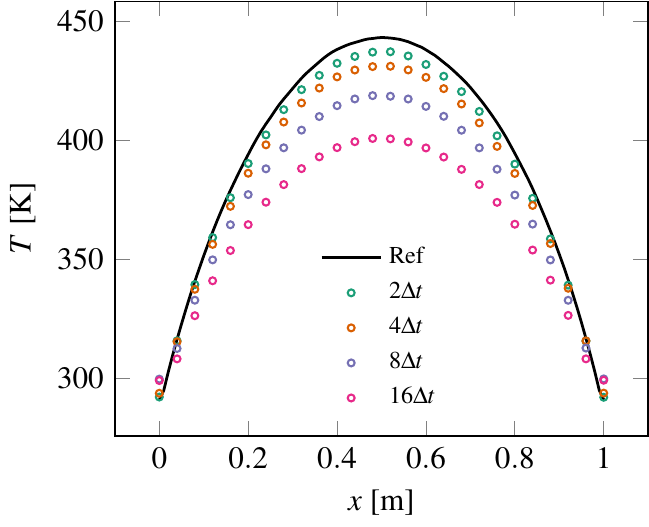}}
  \caption{Comparison of temperature using ED-SP-BGK and SP-BGK as well as different time steps $\Delta t$ for Couette flow with $Kn=0.01$.}\label{fig:couettekn001}
\end{figure}

\begin{figure}\centering
  \subfloat[ED-SPBGK]{\includegraphics{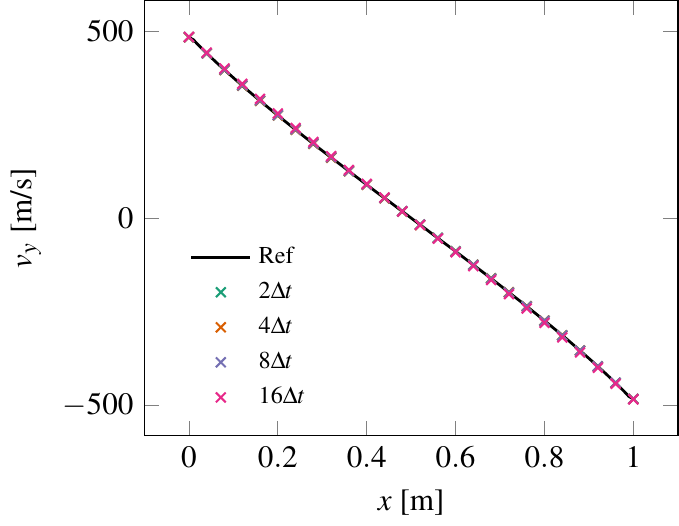}}\\
  \subfloat[SPBGK]{\includegraphics{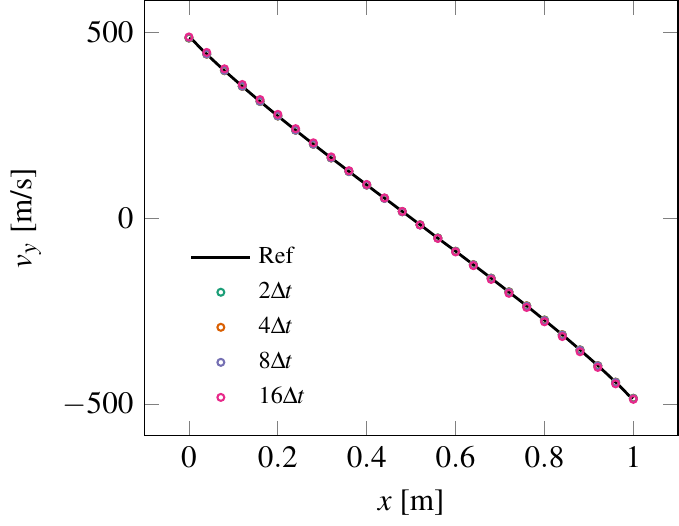}}
  \caption{Comparison of velocity $v_y$ using ED-SP-BGK and SP-BGK as well as different time steps $\Delta t$ for Couette flow with $Kn=0.01$.}\label{fig:couettekn001velo}
\end{figure}

In the transition regime with $Kn=0.01$, the performance of the SP-BGK method becomes worse for larger time step sizes as depicted in \cref{fig:couettekn001}. The non-dimensional time step reaches $\nu\Delta t\approx2$  for the largest time step size of $16\Delta t$. The ED-SP-BGK method still shows good agreements with the reference solution. The time step is bounded by the CFL condition.
The velocity plot depicted in \cref{fig:couettekn001velo} again shows good agreement for both methods.
\subsubsection{Kn=0.001}
\begin{figure}\centering
  \subfloat[ED-SPBGK]{\includegraphics{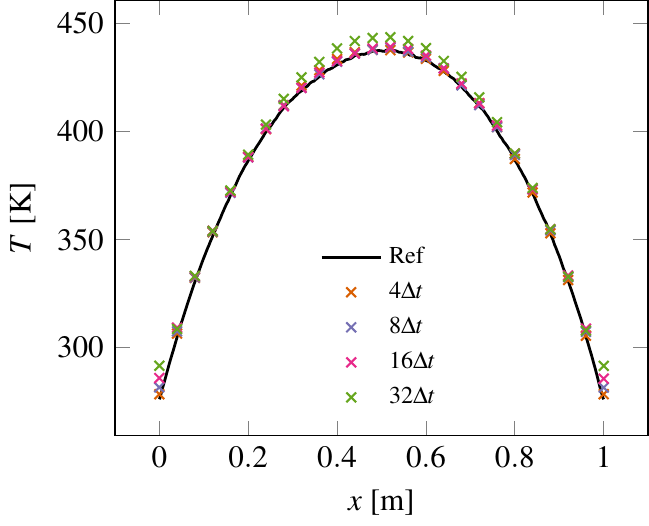}}\\
  \subfloat[SPBGK]{\includegraphics{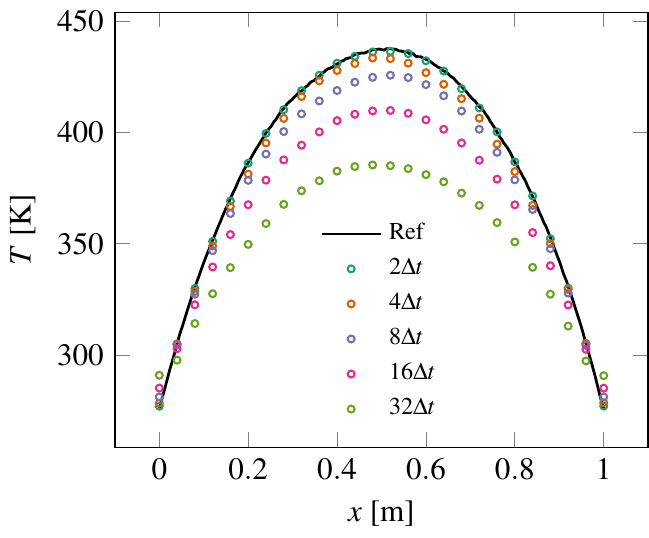}}
  \caption{Comparison of temperature using ED-SP-BGK and SP-BGK as well as different time steps $\Delta t$ for Couette flow with $Kn=0.001$.}\label{fig:couettekn0001}
\end{figure}

\begin{figure}\centering
  \subfloat[ED-SPBGK]{\includegraphics{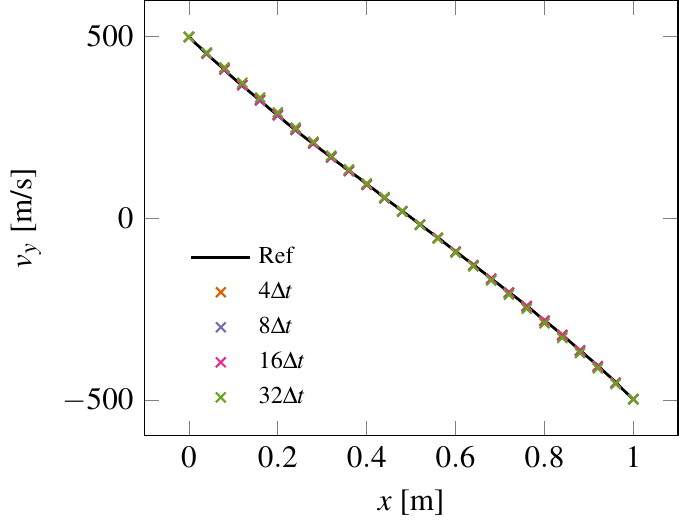}}\\
  \subfloat[SPBGK]{\includegraphics{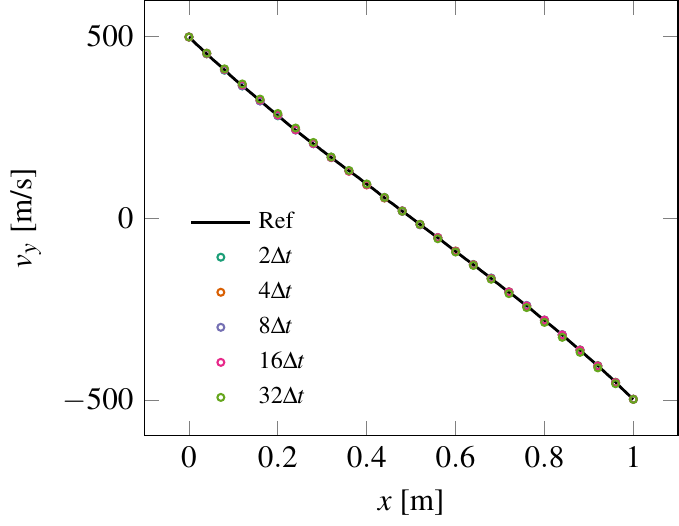}}
  \caption{Comparison of velocity $v_y$ using ED-SPBGK and SP-BGK as well as different time steps $\Delta t$ for Couette flow with $Kn=0.001$.}\label{fig:couettekn0001velo}
\end{figure}

The last test case enters the continuum range with $Kn=0.001$. Here the required time step size to accurately resolve the solution becomes prohibitively small for the SP-BGK method, as shown in \cref{fig:couettekn0001}. We get $\nu\Delta t\approx5$ corresponding to the time step size of $32\Delta t$, where again the ED-SP-BGK scheme performs reasonably well.
For the velocity plots \cref{fig:couettekn0001velo}, the picture is the same as before. Both methods can represent the velocity very well.
\section{Conclusion}
Particle methods comprise an attractive class of algorithms, which are typically employed for rarefied gas simulations. Despite their popularity and robustness, their numerical accuracy and convergence-order hardly go beyond first-order. Multi-scale flow phenomena with realistic rarefaction regimes, however, require high-order schemes and thus entail fundamental improvements on numerical aspects of the stochastic particle methods. This study enhances SP-BGK schemes by introducing a consistent second-order treatment of the BGK relaxation, using the exponential differencing method. The structure of the devised ED-SP-BGK method allows for a straightforward implementation and minimal overhead with respect to the traditional SP-BGK method, as the particle weights remain non-negative irrespective of the spatio-temporal scales.
Besides correct asymptotic limits of the proposed ED-SP-BGK scheme in both free-molecular and continuum limits, its performance was investigated in a few standard settings covering a large variation of the Knudsen number. It is demonstrated that the devised scheme performs better than the SP-BGK method in all considered scenarios including both rarefied and continuum regimes. Extension of the scheme for more complex gas flow scenarios besides improved treatment of the boundary conditions will be pursued in the follow-up studies. 

\section*{Acknowledgments}
This project has received funding from the European Research Council (ERC) under the European Union’s Horizon 2020 research and innovation programme (grant agreement No. 899981 MEDUSA).

\appendix 
\section{Time dependent distribution function in the continuum limit}
\label{sec:append}
This analysis is based on the procedure given by \citet{guo2013discrete}. Let us consider Eq.~\eqref{eq:finblin}
\begin{eqnarray}
&f&(\mathbf v, \mathbf x +\mathbf v \Delta t, t+\Delta t)= \nonumber \\
&&\frac{1-e^{-\nu\Delta t}}{\nu \Delta t} \tilde{f}^*(\mathbf v, \mathbf x +\mathbf v \Delta t, t+\Delta t) \nonumber \\
&+& \left(1-\frac{1-e^{-\nu\Delta t}}{\nu \Delta t}\right)f^M (\mathbf v, \mathbf x +\mathbf v \Delta t, t+\Delta t), \label{eq:app1}
\end{eqnarray}
 where it is assumed $f^t=f^M$. In addition, state $(\mathbf v, \mathbf x +\mathbf v \Delta t, t+\Delta t)$ is abbreviated as $(t+\Delta t)$ and state $(\mathbf v, \mathbf x, t)$ as $(t)$.
It is known that $\tilde{f}^*(\mathbf v, \mathbf x +\mathbf v \Delta t, t+\Delta t)$ results from the movement of the particles along the trajectories of $\hat{f}^*(\mathbf v, \mathbf x , t)$ which corresponds to the advection step $\tilde{f}^*(\mathbf v, \mathbf x +\mathbf v \Delta t, t+\Delta t) = \hat{f}^*(\mathbf v, \mathbf x , t) - \mathbf v \Delta t\nabla\hat{f}^*(\mathbf v, \mathbf x , t)$. Substitung this into \eqref{eq:app1} and using \eqref{eq:finblin} yields
\begin{eqnarray}
&f&(t+\Delta t)=\frac{1-e^{-\nu\Delta t}}{\nu \Delta t} \left[\frac{\nu\Delta te^{-\nu\Delta t}}{1-e^{-\nu\Delta t}}(f(t)-\mathbf v\Delta t\nabla f(t) )\right. \nonumber \\
&+&\left. \left(1- \frac{\nu\Delta te^{-\nu\Delta t}}{1-e^{-\nu\Delta t}}\right)(f^M(t)-\mathbf v\Delta t\nabla f^M(t) )\right] \nonumber \\
&+& \left(1-\frac{1-e^{-\nu\Delta t}}{\nu \Delta t}\right)f^M (t+\Delta t)
\label{eq:app2}
\end{eqnarray}
Let $f^M (t+\Delta t)\approx f^M(t)+ \Delta t \partial_t f^M(t)$, which is justified in the continuum limit as discussed in \citet{guo2013discrete}, and suppose that $f(t)$ in the continuum limit can be approximated by the Chapman-Enskog expansion
\begin{equation}
f(t)\approx f^M(t) -\tau D_t f^M(t) + \mathcal{O}(D_t^2)
\end{equation}
with $D_t= (\partial_t + \mathbf v \nabla)$, Eq.~\eqref{eq:app2} leads to
\begin{eqnarray}
&f&(t+\Delta t)=e^{-\nu\Delta t}(f^M(t)-\tau D_t f^M(t))  \nonumber \\
&+& \left(\frac{1-e^{-\nu\Delta t}}{\nu\Delta t}-e^{-\nu\Delta t}\right)f^M(t) \nonumber \\
&+& \left(1-\frac{e^{-\nu\Delta t}}{\nu\Delta t}\right) (f^M(t)+\Delta t \partial_t f^M(t)) \nonumber \\
&-& \mathbf v \Delta t\left[e^{-\nu\Delta t}\nabla f^M(t)+ \left(\frac{1-e^{-\nu\Delta t}}{\nu\Delta t}-e^{-\nu\Delta t}\right)\nabla f^M(t)\right] \nonumber \\
&+&\mathcal{O}(\partial^2)
\end{eqnarray}
Further rearrangements of the prefactors yields 
\begin{eqnarray}
f(t+\Delta t)&=&f^M(t)-\tau D_t f^M(t) \nonumber \\
&+&\Delta t\partial_t f^M(t)+\mathcal{O}(\partial^2)
\end{eqnarray}
which is the time-dependent Chapman-Enskog Navier-Stokes distribution~\cite{guo2013discrete}.
\bibliography{mybibfile}
\end{document}